\documentclass[prl,twocolumn,superscriptaddress,showpacs,floatfix,longbibliography]{revtex4-1}
\usepackage{mathrsfs,braket}
\usepackage{amssymb, amsbsy, amsmath, latexsym, dsfont, array, layout, graphicx, mathrsfs, color, ulem, bm, hyperref}
\usepackage[dvipsnames,table]{xcolor} 
\usepackage{cancel}
\hypersetup{colorlinks, linkcolor={blue!90!black}, citecolor={blue!90!black}, urlcolor={blue!90!black} }

\newcommand{\one}{\mathds{1}}

\begin{document}

\title{Observation of non-Hermitian edge burst in quantum dynamics}

\author{Lei Xiao} \thanks{These authors contributed equally to this work.}
\affiliation{Beijing Computational Science Research Center, Beijing 100084, China}

\author{Wen-Tan Xue} \thanks{These authors contributed equally to this work.}
\affiliation{Institute for Advanced Study, Tsinghua University, Beijing, 100084, China}

\author{Fei Song}
\affiliation{Institute for Advanced Study, Tsinghua University, Beijing, 100084, China}

\author{Yu-Min Hu}
\affiliation{Institute for Advanced Study, Tsinghua University, Beijing, 100084, China}

\author{Wei Yi}\email{wyiz@ustc.edu.cn}
\affiliation{CAS Key Laboratory of Quantum Information, University of Science and Technology of China, Hefei 230026, China}
\affiliation{CAS Center For Excellence in Quantum Information and Quantum Physics, Hefei 230026, China}

\author{Zhong Wang}\email{wangzhongemail@tsinghua.edu.cn}
\affiliation{Institute for Advanced Study, Tsinghua University, Beijing, 100084, China}

\author{Peng Xue}\email{gnep.eux@gmail.com}
\affiliation{Beijing Computational Science Research Center, Beijing 100084, China}

\begin{abstract}

The non-Hermitian skin effect, by which the eigenstates of Hamiltonian are predominantly localized at the boundary, has revealed a strong sensitivity of non-Hermitian systems to the boundary condition. Here we experimentally observe a striking boundary-induced dynamical phenomenon known as the non-Hermitian edge burst, which is characterized by a sharp boundary accumulation of loss in non-Hermitian time evolutions. In contrast to the eigenstate localization, the edge burst represents a generic non-Hermitian dynamical phenomenon that occurs in real time. Our experiment, based on photonic quantum walks, not only confirms the prediction of the phenomenon, but also unveils its complete space-time dynamics.  Our observation of edge burst paves the way for studying the rich real-time dynamics in non-Hermitian topological systems.
\end{abstract}
\maketitle

\begin{figure*}
\centering
\includegraphics[width=0.85\textwidth]{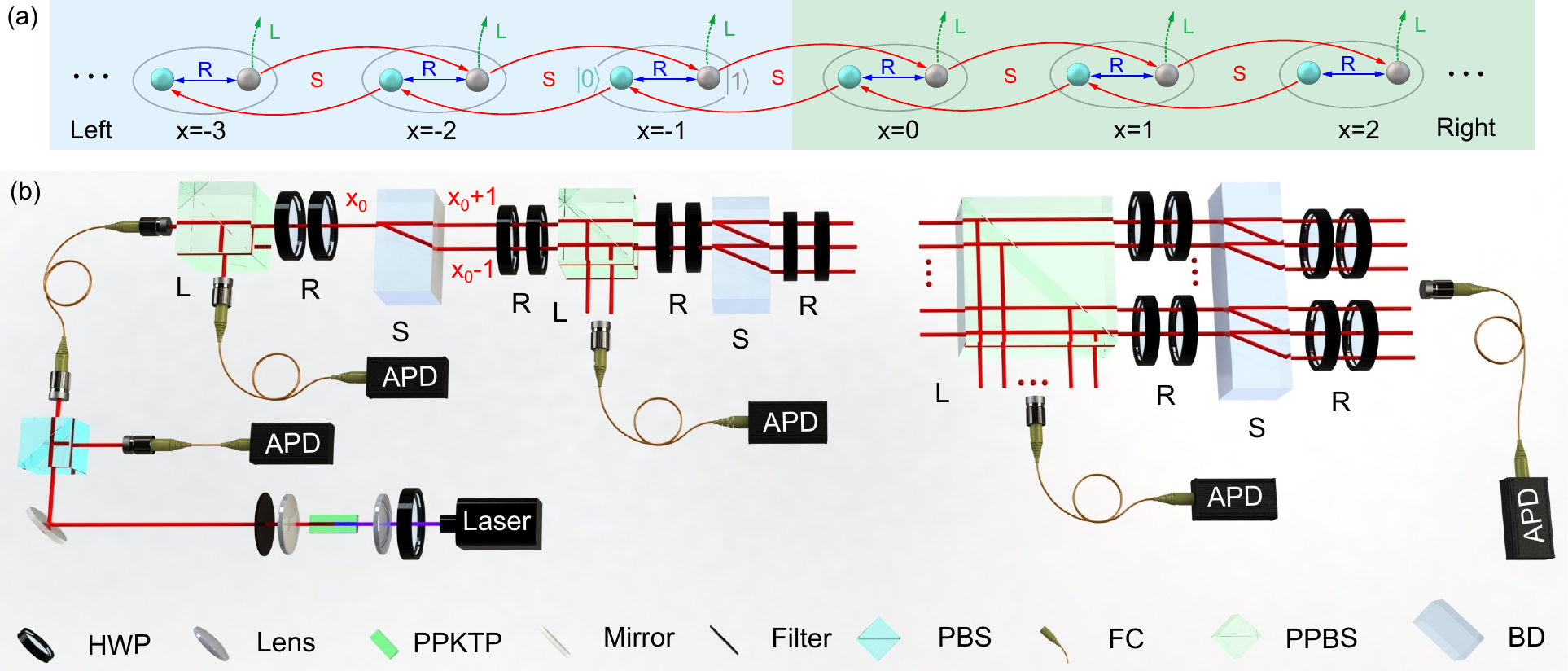}
\caption{Experimental implementation. (a) The domain-wall geometry of the
non-Hermitian quantum walk. The operations of $S, R, L$ contained in $U$ are pictorially shown.  (b) Experimental setup. Photon pairs are created by the spontaneous parametric down conversion process in a type-II cut PPKTP crystal. One of the photon is injected into the quantum-walk interferometric network, and the other is used as the trigger. The walker photon passes the polarizing beam splitter (PBS) and the half-wave plate (HWP), so that its polarization is prepared in the coin state $\ket{0}$. It then undertakes the quantum walk through the network containing partially polarizing beam splitters (PPBSs), HWPs, beam displacers (BDs). Finally, avalanche photodiodes (APDs) are used to detect the walker photons that coincide with the trigger photons.}
\label{fig1}
\end{figure*}

\begin{figure*}
\includegraphics[width=\textwidth]{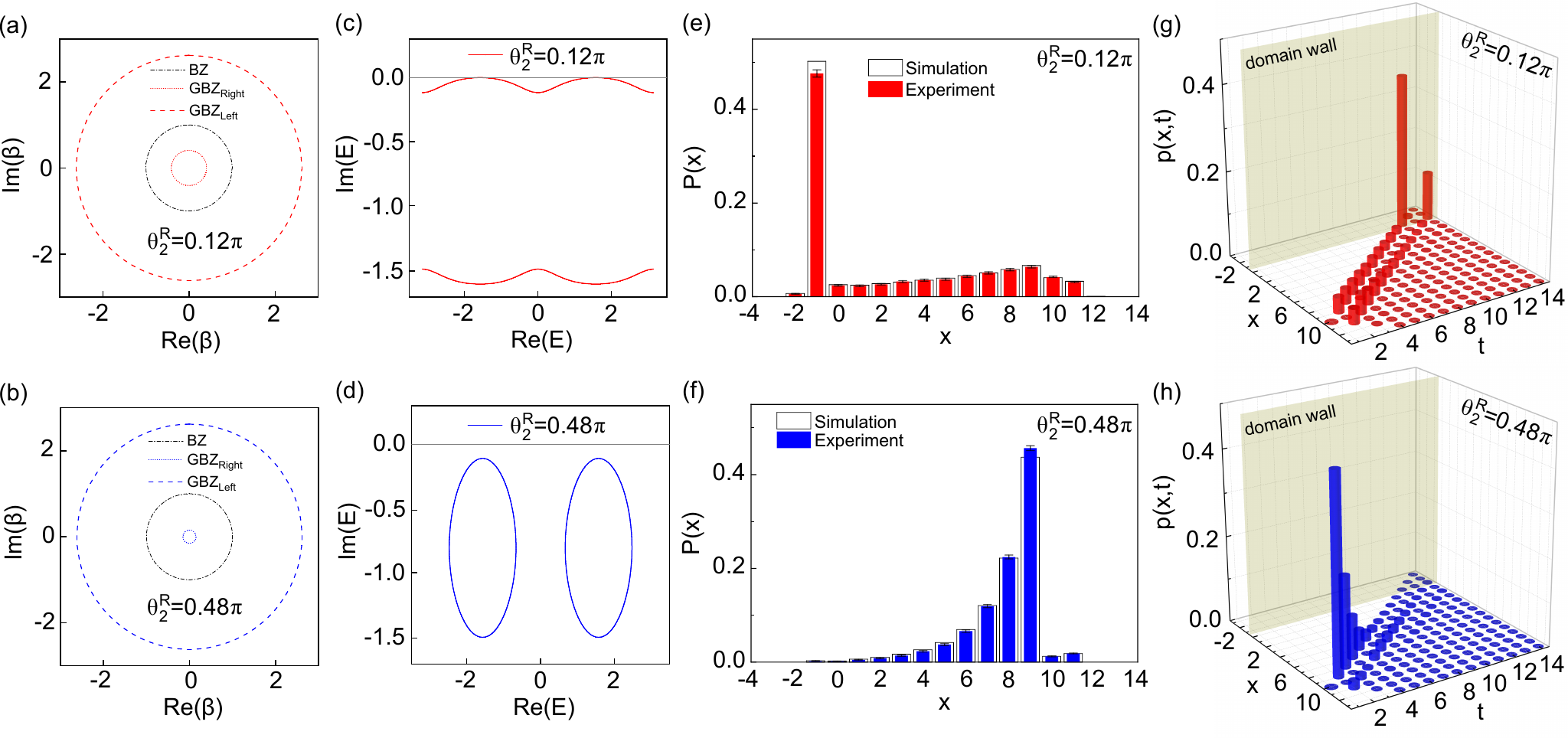}
\caption{Edge burst in non-Hermitian quantum walks. The fixed parameters are $\theta_{1,2}^L=0.85\pi$, $\theta_1^R=0.12\pi$ and $\gamma=0.8$. (a)(b) Brillouin zone (BZ) and generalized Brillouin zone (GBZ) for $\theta_2^R=0.12\pi$ and $\theta_2^R=0.48\pi$. (c)(d) Energy spectra (for the right region in which the walker is initialized) under the periodic boundary condition (PBC) for two indicated values of $\theta_2^R$. (e)(f) Experimentally measured $P(x)$ of a $14$-step non-Hermitian quantum walk with the initial state $\ket{x_0=10}\otimes\ket{0}$. (g)(h) The space-time-resolved loss probability $p(x,t)$ for the two values of $\theta_2^R$. Error bars represent the statistical uncertainty under the assumption of Poissonian statistics.
}
\label{fig2}
\end{figure*}

\begin{figure*}
\includegraphics[width=0.85\textwidth]{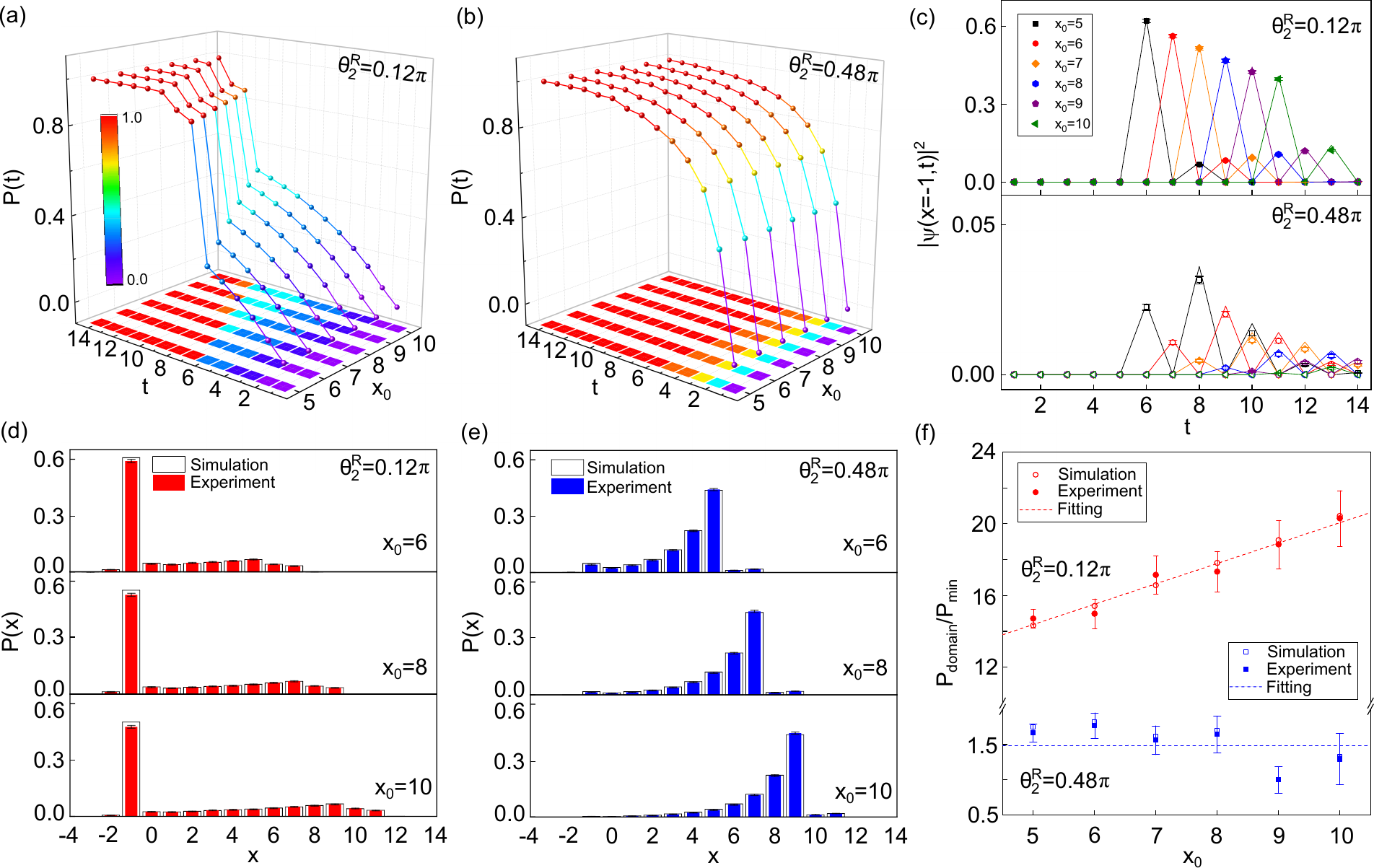}
\caption{(a)(b) Experimentally measured time-dependent total loss probability $P(t)$ for different starting positions $x_0=5,6,7,8,9,10$, respectively. For (a), $\theta_2^R=0.12\pi$; for (b), $\theta_2^R=0.48\pi$. Other parameters are the same as those in Fig.~\ref{fig2}. (c) The measured survival probability at the domain wall, $|\psi(x=-1,t)|^2$, for different starting positions.  $\theta_2^R=0.12\pi$ (upper panel) and $0.48\pi$ (lower panel). (d) (e) Experimentally measured $P(x)$ ($T=14$) for $x_0=6,8,10$; $\theta_2^R=0.12\pi$ for (d) and $0.48\pi$ for (e). (f) The measured relative height $P_{\text{domain}}/P_{\text{min}}$ versus $x_0$. Hollow and solid symbols represent numerically evaluated results and experimental data, respectively. Dashed lines are obtained by numerical fitting experimental data.
}
\label{fig3}
\end{figure*}

Non-Hermitian physics has attracted increasing attention in a vast variety of contexts ranging from classical waves to open quantum systems \cite{Ashida,BBK21}. Intriguingly, the spatial boundary plays a much more dramatic role in non-Hermitian systems than in Hermitian ones. In particular, for certain non-Hermitian systems, the eigenstates concentrate predominantly at the boundary, which is known as the non-Hermitian skin effect (NHSE)~\cite{YW18,YSW18,KEB+18,AVT18,LT19,SYW19,OKS+20,XDW+20,HHI+20,Ma,GBW+20,ZYF22}. Among many other consequences, it implies a fundamental revision of the principle of bulk-boundary correspondence ~\cite{XDW+20,HHI+20}.

Whereas the NHSE has revealed intriguing static properties such as novel behaviors of eigenstates and energy spectra, in this work we unveil a striking dynamic boundary effect in non-Hermitian systems. We experimentally observe that in a class of lossy quantum walk of single photons, the loss rate is drastically enhanced at the boundary. Specifically, for a lossy particle initially located at a position far from the boundary of a lattice system, the space-resolved loss has a surprisingly high boundary peak, in sharp contrast to the common expectation that the particle loss should decay away from the initial position. Remarkably, the relative height of the edge peak even grows as the distance between the initial position and boundary increases. This striking phenomenon, dubbed non-Hermitian edge burst, has been predicted in recent theories \cite{XHS+22,WLZ21}.

Since both the NHSE and edge burst involve boundary localization, it is tempting to attribute the latter to the former. However, it turns out that NHSE does not guarantee the emergence of edge burst. Closing the gap of the imaginary part of energy spectrum (i.e., the imaginary gap or dissipative gap) is the other necessary condition, which highlights the rich implication of spectral profile and topology in non-Hermitian systems~\cite{ZYF22,OKS+20}. At a deeper level, a novel dynamic bulk-edge scaling relation has been suggested as the origin of edge burst \cite{XHS+22}. Thus, the edge burst signifies an unprecedented interplay between non-Hermitian topological physics and non-Hermitian dynamical phenomena.

{\it Lossy quantum walk.---}To study the non-Hermitian edge burst, we design a one-dimensional quantum walk \cite{ADZ93,FG98,W01,C09} with the Floquet operator \begin{equation}
U=R\left(\frac{\theta_2}{2}\right)SR\left(\frac{\theta_1}{2}\right)L(\gamma).
\label{U}
\end{equation}
The shift operator $S=\sum_{x}|x-1\rangle\langle x|\otimes| 0\rangle\langle 0|+| x+1\rangle\langle x|\otimes| 1\rangle\langle 1|$, so that the walker's position is shifted from the site $x$ to $x-1$ or $x+1$ according to the coin state $\ket{0}$ or $\ket{1}$. The coin state is rotated along the $y$ axis by $R(\theta) =\one_\text{w} \otimes e^{-i \theta \sigma_{y}}$, where $\one_\text{w}=\sum_x\ket{x}\bra{x}$ is the identity operator. The operator $L(\gamma)=\one_\text{w} \otimes \begin{pmatrix}
	1 & 0 \\
	0 & e^{-2\gamma}
	\end{pmatrix}$ generates a state-selective loss. For our photonic platform, it is more convenient to create a domain wall instead of an open boundary [see Fig.~\ref{fig1}(a)].  The left (L) and right (R) regions are characterized by coin parameters $\theta_{1,2}^L$ and $\theta_{1,2}^R$, respectively. The dynamics of the non-Hermitian quantum walk follows
\begin{equation} \label{t}
|\psi(t)\rangle=U^t|\psi(0)\rangle,
\end{equation}
where $\ket{\psi(0)}$ is the initial state and $t$ is the integer discrete time.  One can also define an effective non-Hermitian Hamiltonian $H_\text{eff}$ by $U=\exp(-iH_\text{eff})$,  which shares the same eigenstates as $U$.

The Floquet operator $U$ defined in Eq.~(\ref{U}) and the associated $H_\text{eff}$ exhibit the NHSE,  which originates from the state-dependent directional hoppings built in the model (akin to Refs. \cite{XDW+20,XDW+21}). In the presence of a domain wall [Fig.~\ref{fig1}(a)], all the eigenstates of $U$ exhibit localization at the domain wall when the non-Hermiticity is nonzero, i.e., $\gamma\neq 0$. Accordingly, the generalized Brillouin zone (GBZ) deviates from the unit circle [see Figs.~\ref{fig2}(a) and (b)] \cite{YW18,Yokomizo2019,DY19}. Here, we focus on two sets of parameters, $\theta_2^R=0.12\pi$ and $\theta_2^R=0.48\pi$,  with other parameters fixed as $\theta_{1,2}^L=0.85\pi$, $\theta_1^R=0.12\pi$, and $\gamma=0.8$. In Figs.~\ref{fig2}(c) and (d), we show the energy spectrum of $H_\text{eff}$, which clearly indicates that the imaginary gap (the gap between $0$ and the maximum imaginary part of the spectrum) is zero for $\theta_2^R=0.12\pi$ but nonzero for $\theta_2^R=0.48\pi$. In fact, the imaginary gap vanishes along the lines $\theta_1=2\pi n\pm\theta_2$ ($n\in \mathbb{Z}$) (see Supplementary Information).

{\it Observation of edge burst.---}In our experiment, a walker is initialized at a site $x_0$, which  evolves under Eq.~\eqref{t} in discrete time steps. The key quantity for edge burst is the probability $P(x)$ that the walker escapes from the position $x$. In practice, one can measure the space-time-resolved loss $p(x,t)$ from $t=1$ to $t=T$, with $T$ being a large integer so that the loss is almost complete. The sum over $t$ then gives
\begin{equation}
P(x)=\sum_{t=1}^{T}p(x,t).
\label{P}
\end{equation}
According to the specific form of loss adopted here, we have
\begin{align}
\label{eq:p}
p(x,t)=(1-e^{-4\gamma})|\bra{1}\otimes\bra{x}\psi(t-1)\rangle|^2.
\end{align}  It may also be written as $p(x,t)=|\bra{1}\otimes\bra{x}M|\psi(t-1)\rangle|^2$ with $M=\one_\text{w}\otimes\begin{pmatrix}
	0 & 0 \\
	0 & \sqrt{1-e^{-4\gamma}} \end{pmatrix}$, which can be implemented by a partial measurement via the PPBS [see Fig. \ref{fig1}(a)] at the time step $t$. We also define a time-dependent total loss probability \begin{equation}
\label{eq:PL}
P(t)=\sum_{t'=1}^t \sum_{x}p(x,t'),
\end{equation} so that the survival probability after a $t$-step evolution is $1-P(t)$. In our quantum-walk platform, $p(x,t)$ can be readily extracted from photon-number measurements (see Methods), and $P(x)$, $P(t)$ can be obtained from Eqs.~\eqref{P}\eqref{eq:PL}.

We implement a $14$-step ($T=14$) quantum walk with initial walker location $x_0=10$. The space-resolved loss probability $P(x)$ is shown in Figs.~\ref{fig2}(e) and (f) for the  aforementioned two sets of parameters. In both (e) ($\theta_2^R=0.12\pi$) and (f)($\theta_2^R=0.48\pi$), we observe that the loss probability initially decays away from $x_0$. Moreover, the $P(x)$ profile is asymmetric around $x_0$, which can be naturally attributed to the NHSE.

The surprising feature is an exceptionally high peak emerging at the domain wall in Fig.~\ref{fig2}(e). Intuitively, one may resort to the NHSE to explain this edge burst.  However, the NHSE is also strong for the parameters of Fig.~\ref{fig2}(f), yet the edge burst is not seen there. Therefore, the origin of edge burst cannot be explained by the NHSE alone. In fact, the imaginary gap plays an essential role here \cite{XHS+22}. The corresponding imaginary gap, shown in Figs.~\ref{fig2}(c) and (d), is zero and nonzero for Figs.~\ref{fig2}(e) and (f), respectively.

To unveil the space-time profile of walker's loss, we plot $p(x,t)$ for the above two sets of parameters. Figs.~\ref{fig2}(g) and (h)  show that the walker propagates almost ballistically with concurrent loss along the trajectory. In the case of edge burst, a large loss peak in $p(x,t)$ emerges when the walker hits the domain wall. It also indicates that the burst occurs around a particular time, before which it is indiscernible.

Furthermore, we vary the initial position $x_0=5,6,7,8,9,10$ and measure the time-dependent loss probability $P(t)$. As shown in Fig.~\ref{fig3}(a), for $\theta_2^R=0.12\pi$ (with edge burst), $P(t)$ suddenly increases near the domain wall. In contrast, in Fig.~\ref{fig3}(b), for $\theta_2^R=0.48\pi$ (without edge burst), $P(t)$ increases steadily with $t$ without sudden change. Similarly, the space-resolved survival probability $|\psi(x=-1,t)|^2$ at the domain wall at each step $t$ behaves differently with and without the edge burst [see Fig.~\ref{fig3}(c)]. The value of $|\psi(x=-1,t)|^2$ is significantly larger in the presence of edge burst.  In Fig.~\ref{fig3}(d), we show that the edge burst remains robust  when the starting position varies. In contrast, when the edge burst is absent, $P(x)$ decays rapidly as $x_0$ moves away from the domain wall [see Fig.~\ref{fig3}(e)].

To further characterize the edge burst, we measure the relative height $P_\text{domain}/P_\text{min}$, where $P_\text{domain}\equiv P(x=-1)$ is the probability that the photon escapes from the domain wall $x=-1$, and $P_\text{min}\equiv \min_{x=-1,\cdots,x_0}\{P(x)\}$ is the minimum of $P(x)$ in the interval between the initial location $x_0$ and the domain wall location $x=-1$. The edge burst is characterized by $P_\text{domain}/P_\text{min}\gg 1$, while its absence means that $P_\text{domain}/P_\text{min}$ is on the order of unity. As shown in Fig.~\ref{fig3}(f), for $\theta_2^R=0.48\pi$, the measured relative height remains close to $1$ as $x_0$ increases. In stark contrast, for $\theta_2^R=0.12\pi$,  the relative height increases with $x_0$ and fits well with a linear relation $P_\text{domain}/P_\text{min}\sim  x_0$. Thus, the relative height grows as the initial walker position moves away from the domain wall. While counterintuitive, this behavior is a consequence of a novel bulk-edge scaling relation \cite{XHS+22}.

{\it Discussions.---}We present the first experimental observation of the non-Hermitian edge burst by using discrete-time non-Hermitian quantum walk of photons. Our experiment not only demonstrates that edge burst originates from the intriguing interplay between two unique non-Hermitian concepts, the NHSE and imaginary gap, but also unveils the real-time dynamics of this phenomenon. The observation of non-Hermitian edge burst paves the way for investigating the real-time dynamics in non-Hermitian topological systems, which remains largely unexplored. From a practical perspective, the edge burst may offer a promising non-Hermitian approach for the on-demand harvesting of light or particles at a prescribed position.

{\bf Methods}

{\it Implementation.---}For the experimental implementation, we adopt the scheme of single-photon discrete-time quantum walks illustrated in Fig.~\ref{fig1}(b).  Photon pairs are created by spontaneous parametric down conversion, where a $20$mm type-II periodically poled potassium titanyl phosphate (PPKTP) crystal is pumped by a $405$nm continuous wave diode laser with the power of $1$mW. One photon serves as a trigger, and the other as a heralded single photon undertaking the quantum walk. The photon polarizations are adopted as the coin state. The walker photon is initialized in the spatial mode $|x_0\rangle$ with the internal state $|0\rangle$, i.e. $|\psi(0)\rangle=|x_0\rangle\otimes|0\rangle$. The localized initial state is prepared by passing the walker photons through a half-wave plate (HWP) and a polarizing beam splitter (PBS).

For the quantum-walk dynamics, the shift operator $S$ is implemented by a beam displacer (BD) whose optical axis is cut in the way so that the vertically polarized photons are directly transmitted and the horizontally polarized photons are laterally displaced into a neighboring mode. The coin rotation $R(\frac{\theta_{1(2)}}{2})$ is realized by two HWPs at $0$ and $\frac{\theta_{1(2)}}{4}$, respectively. The loss operator $L(\gamma)$ is realized by a partially polarizing beam splitter (PPBS),  which completely transmits the coin state $\ket{0}$ but reflects the coin state $\ket{1}$ with a probability $e^{-4\gamma}$. At last, avalanche photodiodes (APDs) are used to detect the walker photons coinciding with the trigger photons. The total number of coincidences is approximately $23000$.

The measurements are based on photon-number counting. The space-time-resolved probability $p(x,t)$ can be calculated from the photon number through
\begin{equation}
p(x,t)=\frac{N(x,t)}{\sum_{x'}N'(x',t)+\sum_{t'=1}^{t}\sum_{x'}N(x',t')},
\end{equation}
where $N(x,t)$ is the number of photons escaping from the position $x$ at the time step $t$, and $N'(x,t)$ is the number of remaining photons at $x$ after a $t$-step evolution.

Finally, the space-resolved survival probability at $x$ can be calculated as
\begin{equation}
|\psi(x,t)|^2=\frac{N'(x,t)}{\sum_{x'}N'(x',t)+\sum_{t'=1}^{t}\sum_{x'}N(x',t')}.
\end{equation}

{\bf Note.} After completing this work, we learned of a related experiment by a team at Southern University of Science and Technology.

{\bf Acknowledgments}

This work has been supported by the National Natural Science Foundation of China (Grant Nos. 92265209, 12025401,  12125405, 11974331 and 12104036.)









\clearpage
\appendix

\pagebreak
\widetext
\begin{center}
\textbf{\large  Supplemental Material for  ``Observation of non-Hermitian edge burst in quantum dynamics''}
\end{center}
\setcounter{equation}{0}
\setcounter{figure}{0}
\setcounter{table}{0}
\makeatletter
\renewcommand{\theequation}{S\arabic{equation}}
\renewcommand{\thefigure}{S\arabic{figure}}
\renewcommand{\bibnumfmt}[1]{[S#1]}

\section{Effective Hamiltonian and generalized Brillouin zone}
In this section, we derive an expression for the effective Hamiltonian $H_\text{eff}$ in momentum space. First, we transform the real-space nonunitary Floquet operator $U$ and its conjugate transpose $U^\dag$ into the momentum-space $U_k$ and $U_k^\dag$:
\begin{align}
&U_k=d_0\sigma_0-id_1\sigma_1-id_2\sigma_2-id_3\sigma_3,\nonumber\\
&U_k^{\dagger}=d_0^{\ast}\sigma_0+id_1^{\ast}\sigma_1+id_2^{\ast}\sigma_2+id_3^{\ast}\sigma_3
\end{align}
where $\sigma_{1,2,3}$ are the Pauli matrices and $\sigma_0$ is the identity matrix, and
\begin{align}
&d_0=e^{-\gamma}(\cosh\gamma \cos k\cos \frac{\theta_1+\theta_2}{2}+i\sinh\gamma\sin k\cos\frac{\theta_1-\theta_2}{2}),\nonumber\\
&d_1=e^{-\gamma}(\cosh\gamma \sin k\sin \frac{\theta_1-\theta_2}{2}+i\sinh\gamma\cos k\sin\frac{\theta_1+\theta_2}{2}),\nonumber\\
&d_2=e^{-\gamma}(\cosh\gamma \cos k\sin \frac{\theta_1+\theta_2}{2}-i\sinh\gamma\sin k\sin\frac{\theta_1-\theta_2}{2}),\nonumber\\
&d_3=e^{-\gamma}(-\cosh\gamma\sin k\cos\frac{\theta_1-\theta_2}{2}+i\sinh \gamma \cos k \cos\frac{\theta_1 + \theta_2}{2}).
\end{align}
Note that the relation $\sqrt{d_0^2+d_1^2+d_2^2+d_3^2}=e^{-\gamma}$ is satisfied. The eigenvalue and eigenvector can be derived from
\begin{align}
U_k|\psi_{\pm}\rangle=\lambda_{\pm}|\psi_{\pm}\rangle,~
U_k^\dagger|\chi_{\pm}\rangle=\lambda^\ast_{\pm}|\chi_{\pm}\rangle.
\end{align}
Straightforward calculations lead to
\begin{equation}
\lambda_\pm=d_0\pm i t_0,~
\lambda^{\ast}_\pm=d_0^{\ast}\mp i t_0^{\ast},
\label{evalue}
\end{equation}
\begin{align}
|\psi_{\pm}\rangle&=\frac{1}{d_1+id_2}\left(\begin{array}{c}
d_3\mp t_0\\
d_1+id_{2}
\end{array}\right),\nonumber\\
\langle \chi_{\pm}|&=\frac{1}{d_1-id_2}(d_3 \mp t_0,d_1-id_{2}),
\label{evector}
\end{align}
where $t_0=\sqrt{e^{-2\gamma}-d_0^2}$. Since the effective Hamiltonian $H_\text{eff}(k)$ is related to $U_k$ through $U_k=e^{-i H_\text{eff} }$, the quasienergy spectrum of $H_\text{eff}(k)$ is
\begin{equation}
E_{\pm}(k)=i\ln \lambda_{\pm}(k)=\pm \arccos(\cosh\gamma \cos k\cos \frac{\theta_1+\theta_2}{2}+i\sinh\gamma\sin k\cos\frac{\theta_1-\theta_2}{2})-i\gamma.
\label{E}
\end{equation}
Specifically, for $\theta_1=\theta_2$, we have
\begin{equation} E_-(k=\pi/2)=-\text{arccos}(i\sinh\gamma)-i\gamma= -\text{arccos}(\sin i\gamma)-i\gamma=-\pi/2, \end{equation}
so that $\text{Im}[E_-(k=\pi/2)]=0$, i.e., the imaginary gap closes at $k=\pi/2$. For $\theta_1=-\theta_2$, the imaginary gap closes at $k=0$ because
\begin{equation} E_+(k=0)=\text{arccos}(\cosh\gamma)-i\gamma= \text{arccos}(\cos(i\gamma))-i\gamma=0.\end{equation}  Thus,  the imaginary gap closes when $\theta_1=2\pi n\pm\theta_2$ ($n\in \mathbb{Z}$).
While the eigenvectors in Eq.~(\ref{evector}) are not orthogonal, one can derive a set of bi-orthonormal eigenvectors  $\{|\tilde{\psi}_{\pm}\rangle,|\tilde{\chi}_{\pm}\rangle\}$:
\begin{align}
|\tilde{\psi}_{\pm}\rangle=&\frac{|\psi_{\pm}\rangle}{\sqrt{\langle\chi_{\pm}|\psi_{\pm}\rangle}}=\frac{1}{\sqrt{2t_0(t_0\mp d_3)}}\left(\begin{array}{c}
d_3\mp t_0\\
d_1+id_{2}
\end{array}\right),\nonumber\\
\langle \tilde{\chi}_{\pm}|=&\frac{\langle \chi_{\pm}|}{\sqrt{\langle\chi_{\pm}|\psi_{\pm}\rangle}}=\frac{1}{\sqrt{2t_0(t_0\mp d_3)}}(d_3 \mp t_0,d_1-id_{2}),
\end{align}
which satisfy
\begin{equation}
\langle \tilde{\chi}_{m}|\tilde{\psi}_{n}\rangle=\delta_{mn},~
\sum_{m=+,-}|\tilde{\psi}_{m}\rangle\langle \tilde{\chi}_{m}|=\one.
\end{equation}
It follows that
\begin{equation}
U_k=\lambda_{+}|\tilde{\psi}_{+}\rangle\langle \tilde{\chi}_{+}|+\lambda_{-}|\tilde{\psi}_{-}\rangle\langle\tilde{\chi}_{-}|,
\end{equation}
and the effective Hamiltonian $H_{\text{eff}}(k)$ can be written as
\begin{equation}
H_\text{eff}=i\ln{\lambda_{+}}|\tilde{\psi}_{+}\rangle\langle \tilde{\chi}_{+}|+i\ln{\lambda_{-}}|\tilde{\psi}_{-}\rangle\langle \tilde{\chi}_{-}|.
\end{equation}

To derive the generalized Brillouin zone (GBZ)~\cite{YW18,Yokomizo2019}, we rewrite the Floquet operator $U$ as
\begin{align}
U=\sum_{x}|x-1\rangle \langle x|\otimes A_0+|x+1\rangle \langle x|\otimes A_1,
\end{align}
where
\begin{align}
A_0=&R_c(\frac{\theta_2}{2})P_0R_c(\frac{\theta_1}{2}) L_c(\gamma),\nonumber\\
A_1=&R_c(\frac{\theta_2}{2})P_1R_c(\frac{\theta_1}{2}) L_c(\gamma),
\end{align}
with $L_c(\gamma)=\left( \begin{array}{@{\,}cc@{\,}}
	1 & 0 \\
	0 & e^{-2\gamma} \\
	\end{array} \right)$, $R_c(\theta)=e^{-i \theta \sigma_y}$, $P_0=|0\rangle \langle 0|$ and $P_1=|1\rangle \langle 1|$. In view of the translational symmetry inside the bulk, the eigenstate $|\varphi\rangle$ of $U$ can be expressed as
\begin{equation}
|\varphi\rangle=\sum_{x,j}\beta_j^x|x\rangle\otimes|\phi_j\rangle_c,
\label{eigenstate}
\end{equation}
where $|\phi_j\rangle_c$ is the coin state and $\beta_{j}$ is the spatial-mode function. Inserting Eq.~(\ref{eigenstate}) into eigen-equation $U|\varphi\rangle=\lambda|\varphi\rangle$, we obtain
\begin{equation}
(A_0\beta+\frac{A_1}{\beta}-\lambda)|\phi\rangle_c=0,
\label{eigeneq}
\end{equation}
which has nontrivial solutions only when
\begin{equation}
\det[A_0\beta+\frac{A_1}{\beta}-\lambda]=0.
\label{det}
\end{equation}
In an explicit form, Eq.~(\ref{det}) is a quadratic equation of $\beta$:
\begin{equation}
[\sin(\frac{\theta_1}{2})\sin(\frac{\theta_2}{2})-e^{2\gamma}\cos(\frac{\theta_1}{2})\cos(\frac{\theta_2}{2})]\beta^2+(\frac{1}{\lambda}+e^{2\gamma}\lambda)\beta+e^{2\gamma}\sin(\frac{\theta_1}{2})\sin(\frac{\theta_2}{2})
-\cos(\frac{\theta_1}{2})\cos(\frac{\theta_2}{2})=0.
\label{quadratic}
\end{equation}
In the thermodynamic limit, the GBZ equation is determined by $|\beta_1(\lambda)|=|\beta_2(\lambda)|$~\cite{YW18,Yokomizo2019}. Thus, we obtain
\begin{equation}
|\beta_1|=|\beta_2|=\sqrt{|\frac{e^{2\gamma}\sin(\frac{\theta_1}{2})\sin(\frac{\theta_2}{2})
-\cos(\frac{\theta_1}{2})\cos(\frac{\theta_2}{2})}{\sin(\frac{\theta_1}{2})\sin(\frac{\theta_2}{2})-e^{2\gamma}\cos(\frac{\theta_1}{2})\cos(\frac{\theta_2}{2})}|}=\sqrt{|\frac{\cosh \gamma\cos\frac{\theta_1+\theta_2}{2}-\sinh\gamma\cos\frac{\theta_2-\theta_1}{2}}{\cosh \gamma\cos\frac{\theta_1+\theta_2}{2}+\sinh\gamma\cos\frac{\theta_2-\theta_1}{2}}|}.
\label{solution}
\end{equation}
Therefore, the GBZ is a circle in the complex plane, as shown in Fig.~2 in the main text. When $|\beta|<1 (|\beta|>1)$, the skin modes are localized at the left (right) edge. According to Eq.~(\ref{solution}), when $\cos\frac{\theta_1+\theta_2}{2}\cos\frac{\theta_2-\theta_1}{2}>0$ $(\cos\frac{\theta_1+\theta_2}{2}\cos\frac{\theta_2-\theta_1}{2}<0)$, the skin modes are localized at the left (right) edge. For the two sets of parameters used in the main article, the skin modes are localized at the domain wall.

\section{Numerical fitting for larger time steps}

In the experiment, we have found that the relative height can be well fitted by $P_\text{domain}/P_\text{min}\sim x_0$. Thus, the relative height grows as $x_0$ increases. In this section, we add numerical simulations with more steps to further demonstrate this behavior.

\begin{figure*}
\centering
\includegraphics[width=0.95\textwidth]{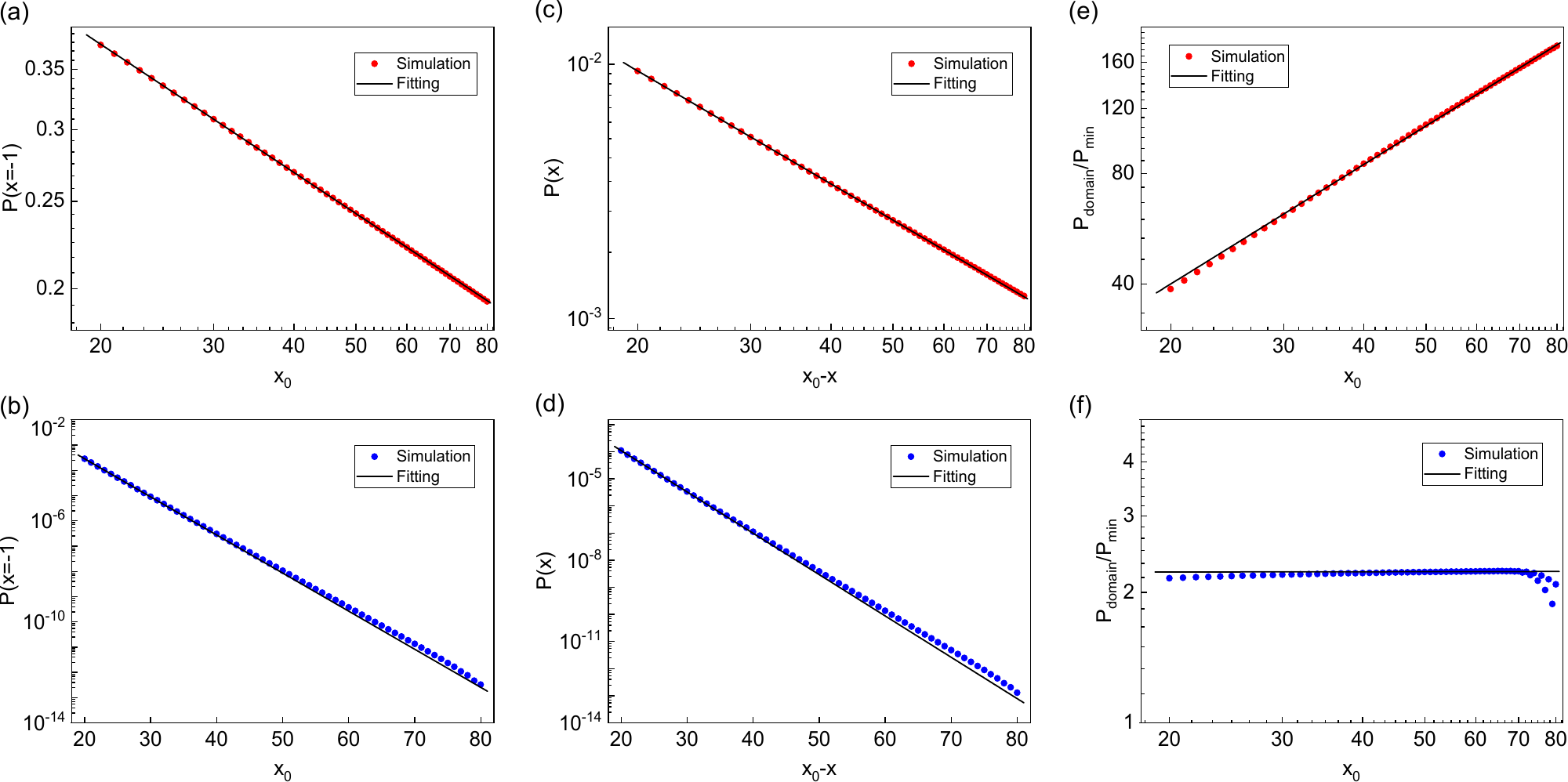}
\caption{ Numerical simulations for the loss probabilities $P(-1)$ (for the domain wall) and $P(x)$ (for the bulk), and the relative height $P_\text{domain}/P_\text{min}$ .  The coin parameters are fixed as $\theta_{1,2}^L=0.85\pi$ and $\theta_1^R=0.12\pi$. For the upper row (red), $\theta_2^R=0.12\pi$, and the edge burst is present; for the lower row (blue),  $\theta_2^R=0.48\pi$, and the edge burst is absent. (a)(b) The loss probability $P(x=-1)$ versus $x_0$.  (c)(d) $P(x)$ versus $x_0-x$.  (e)(f)  The relative height $P_\text{domain}/P_\text{min}$. The dots are from \ numerical simulations, and the black solid lines are the fitting results.}
\label{figS1}
\end{figure*}

As illustrated in Fig.~\ref{figS1}, we fit the loss probability $P(x=-1)$ at the domain wall, $P(x)$ in bulk, and the relative height $P_\text{domain}/P_\text{min}$. The results show that when the edge burst exists [Fig.~\ref{figS1}(a,c,e)], both $P(x=-1)$ and $P(x)$ follow power laws: $P(x=-1)\sim x_0^{-\alpha_d}$ and $P(x)\sim (x_0-x)^{-\alpha_b}$, with certain $\alpha_d$ and $\alpha_b$. The fitting for the relative height is $P_\text{domain}/P_\text{min}\sim x_0^{1.0802}$, which is close to the $P_\text{domain}/P_\text{min}\sim x_0$ behavior predicted by theory and supported by our experiment. Notably, the fitting for $\alpha_{d,b}$ are $\alpha_d=0.4717$ and $\alpha_b=1.4751$, so that $\alpha_b-\alpha_d=1.0034$, which agrees well with the predicted bulk-edge scaling relation in Ref.~\cite{XHS+22}.

When the edge burst is absent [Fig.~\ref{figS1}(b,d,f)], the fitting turns out to be exponential: $P(x=-1)\sim \beta_d^{x_0-x}$ and $P(x)\sim \beta_b^{x_0-x}$, with certain $\beta_b$ and $\beta_d$ that are approximately equal. The relative height $P_\text{domain}/P_\text{min}$ is almost constant as $x_0$ varies.

\end{document}